\documentclass{aipproc}

\layoutstyle{6x9}

\usepackage{graphicx}
\begin{document}
\title{Refractive Distortions of Two-Particle Correlations}


\author{Scott Pratt}{
address={Department of Physics and Astronomy,
Michigan State University\\ East Lansing, Michigan 48824}
email={pratts@pa.msu.edu}
}

\begin{abstract}
Using optical model calculations it has recently been shown that refractive phenomena from the collective mean field can significantly alter the sizes inferred from two-pion correlations. We demonstrate that such effects can be accounted for in classical calculations if mean field effects are included.
\end{abstract}

\date{\today}

\maketitle

The six-dimensional two-pion correlation function $C({\bf P},{\bf Q})$, measured as a function of the total and relative momenta, has provided crucial insight into the space-time development of heavy-ion reactions \cite{Lisa:2005dd}. Recently, the refractive effects of particles traversing the mean field were calculated by symmetrizing quantum waves in the frame work a time-independent complex optical potential \cite{Cramer:2004ih,Miller:2005ji}. If the pions left a region of lower mass, it was found that the lensing effect of the potential was to distort the extracted source dimensions by a few tens of percent. The effect was analagous to lensing effects from Coulomb mean fields, which were shown to explain different apparent source sizes for positive and negative pions at AGS energies \cite{Barz:1998ce,Barz:1996gr}.

The refractive corrections studied in Reference \cite{Cramer:2004ih,Miller:2005ji} allowed for a more physical interpretation of correlation data from RHIC. When ignoring the corrections it appears that the fireball grows to a transverse (to the beam) radius of $\approx 13$ fm, expanding at a speed $\approx 0.7c$, and rapidly disintegrates at a time of $\approx 10$ fm/$c$ \cite{Retiere:2003kf}. The puzzling aspect of this picture is that the fireball surface must expand at a speed of 0.7$c$ from the initial collision to grow from its initial size of 6 fm to 13 fm in 10 fm/$c$, not allowing any time for the matter to accelerate. Correlation analyses typically provide three dimensions: $R_{\rm long}$, the longitudinal size defined parallel to the beam, $R_{\rm out}$, the dimension of the phase space packet perpendicular to the beam and outward along the direction of the pair's momentum, and $R_{\rm side}$, the sideward dimension which is perpendicular to both the pair momentum and the beam axis. The apparent sideward dimension in \cite{Cramer:2004ih,Miller:2005ji} was shown to be increased by the refractive effects of the potential, which would suggest the true radius of the fireball was somewhat smaller than the 13 fm previously believed. Reducing the size by one or two fm would provide a more physically plausible picture of the reaction's evolution. 

The calculations in Ref. \cite{Cramer:2004ih,Miller:2005ji} involved solving for single-particle outgoing wave functions in the presence of complex optical potentials, then using the symmetrized product as a correlation weight. Quantum calculations such as these have some drawbacks. First, implementation becomes complictated if one were to account for time-varying potentials, or potentials without the spherical and boost symmetries assumed in \cite{Cramer:2004ih,Miller:2005ji}. Secondly, finding expressions for the mean field, especially the imaginary part of the optical potential, would be quite involved. Finally, it is difficult to extend such calculations beyond the case of identical particles as one then needs to consider the quantum three-body problem. For these reasons, we wish to study whether the effects can be calculated by considering classical trajectories through the mean field, then using the asymptotic phase space density to generate correlations. Given that the inferred diameters of the RHIC fireballs are often near 25 fm, one might expect that classical considerations could be valid except at very low $p_t$. 

For non-interacting identical particles, two-particle probabilities can be calculated in terms of the outgoing phase space density \cite{Lisa:2005dd}. 
\begin{eqnarray}
\label{eq:cf}
C({\bf P},{\bf Q})&=&1+\int d^3r {\mathcal S}_{{\bf P}}({\bf r})
\cos({\bf Q}'\cdot{\bf r}),\\
\nonumber
{\mathcal S}_{\bf P}({\bf r})&=&
\frac{\int d^3r'_a d^3r'_b
f({\bf P}'/2,{\bf r'_a},t') f({\bf P}'/2,{\bf r}'_b,t')
\delta({\bf r}'_a-{\bf r}'_b-{\bf r})}
{\int d^3r'_a d^3r'_b
f({\bf P}'/2,{\bf r'_a},t')f({\bf P}'/2,{\bf r}'_b,t')},
\end{eqnarray}
where the primes denote the positions measured in the pair frame, and ${\bf Q}'={\bf p}'_a-{\bf p}'_b$ is the relative momentum in that frame. Since ${\mathcal S}$ describes the separations of particles with the same velocity, the expression is independent of $t'$ once it satistifies the criteria of being beyond the point at which particles interact with the remainder of the system, including mean-field interactions.

An alternative approach is to treat interaction with the mean field separately,  define the source function so that it describes the points at which particles had their last interaction other than through the mean field, then use the outgoing wave function, which is a solution to the equations of motion using the mean field, to describe the evolution from $x$ to the asymptotic momentum state \cite{Cramer:2004ih,Miller:2005ji,Barz:1998ce}. This involves replacing the phase factor used to describe the evolution from $x$ to its asymptotic state ${\bf p}$,
\begin{equation}
e^{i(p_a-p_b)\cdot x}\rightarrow \phi^*(p_a,x)\phi(p_b,x).
\end{equation}
The two-particle probability is then,
\begin{eqnarray}
\label{eq:cm}
\frac{dN}{d^3p_ad^3p_b}&=&
\int d^4x ~\tilde{s}(p_a,x)|\phi(p_a,x)|^2\int d^4x ~\tilde{s}(p_b,x) |\phi(p_b,x)|^2\\
\nonumber
&+&\left|\int d^4x ~\tilde{s}(P/2,x)
\phi^*(p_a,x)\phi(p_b,x)\right|^2.
\end{eqnarray}
The correlation function is the ratio of the two terms. Here, $\phi$ is the outgoing single-particle wave function describing evolution through the mean field. The source functions $\tilde{s}(p,x)$ now refer to the points where a particle had its last non-mean-field interaction.

Both approaches should be equivalent if the classical trajectories would correctly describe propagation through the mean field. We present an "apples-to-apples" comparison of the two approaches using the same complex optical potential,
\begin{equation}
\label{eq:optical}
U(r)=(U_R+iU_I)\frac{1}{e^{(r-R)/a}+1},
\end{equation}
where $a$ is the diffuseness parameter. If $a=0$, the potential is a step function. We solve for the distortion to the apparent sideward radii for particles which originate isotropically from points a radius $r_0$ from the center of the cylinder. For identical particles, radii can be determined from correlations using the expression \cite{Lisa:2005dd},
\begin{eqnarray}
\label{eq:hbtmoments}
\langle x_ix_j\rangle&=&
\left.\frac{1}{2}\frac{d^2C({\bf P},{\bf Q})}{dQ_idQ_j}\right|_{{\bf Q}=0},
\end{eqnarray}
where the radii represent the dimensions of the asymptotic phase space density, not the dimensions of the emission points. Using Eq. (\ref{eq:cm}) which gives the correlation function in terms of the distributions points for last collisions convoluted with outgoing wave functions, one can then apply Eq. (\ref{eq:hbtmoments}) to write an expression for the variance of the sideward size,
\begin{equation}
R_{\rm side}^2\equiv\langle y^2\rangle
=\frac{\int d^4x \tilde{s}(p,x) \frac{d}{dp_y}\phi^*({\bf p},x)
\frac{d}{dp_y}\phi({\bf p},x)}
{\int d^4x \tilde{s}(p,x)\phi^*({\bf p},x)\phi({\bf p},x)},
\end{equation}
where the asymptotic momentum ${\bf p}$ moves along the $x$ axis and $\phi$ are asymptotic outgoing wave functions.

\begin{figure}
\centerline{\includegraphics[width=0.5\textwidth]{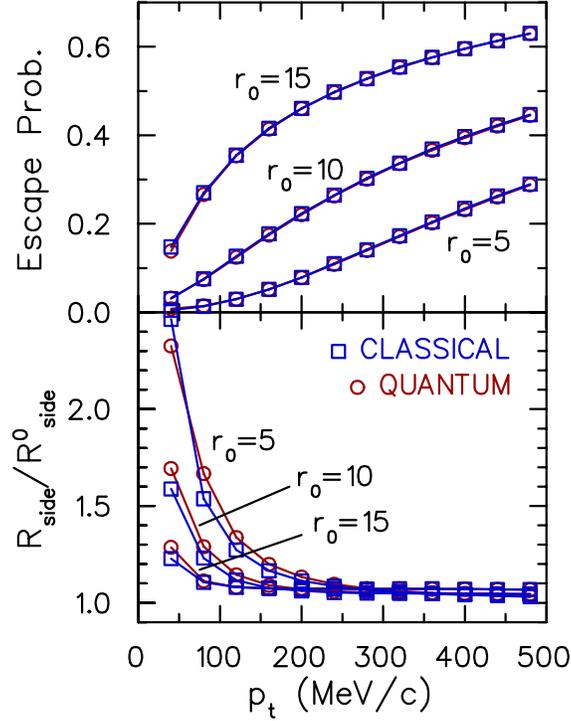}}
\caption{\label{fig:qclass}
The ratio of the apparent sideward dimension to the dimension without mean field is shown as a function of $p_t$ for the optical potential described in the text. Distortions are calculated for both quantum calculations (circles) and classical trajectory calculations (squares). The source function describing the final collision points was confined to an intial radius $r_0$=5,10 or 15 fm. The upper panel shows the probability that such particles escape without being absorbed due to the imaginary part of the optical potential. The two approaches agree within a few percent.}
\end{figure}

Calculations were performed for $\tilde{s}$ being independent of the direction of ${\bf p}$ with emission points being confined to a radius of $r_0$. The potential parameters were $U_R=m_{\rm med}^2-m_{\rm vac}^2$ with the in-medium mass $m_{\rm med}=50$ MeV$/c^2$ and $U_I=m_\pi\cdot\Gamma_0$ with $\Gamma_0=100$ MeV. This corresponds to a classical decay rate of $\Gamma=\Gamma_0 m_\pi/E$. The fact that the decay rate falls $\sim 1/E$ is characteristic of a scalar form for the optical potential. The lower panel of Fig. \ref{fig:qclass} shows the distortion of $R_{\rm side}$, defined as the ratio of $R_{\rm side}$ with the potential to $R_{\rm side}$ with $U=0$, as a function of $p_t$. The distance scales for the potential were $R=10$ fm and $a=3$ fm. Results are shown for three different values of $r_0$: 5 fm, 10 fm and 15 fm. The distortions rise at low $p_t$ with stronger distortions for small $r_0$. Also displayed in Fig. \ref{fig:qclass} are escape probabilities. Quantum mechanically, the probability that a particle escapes the fireball is
\begin{equation}
P_{\rm escape}=\frac{\int d^4x \tilde{s}(p,x) |\phi({\bf p},{\bf r})|^2}
{\int d^4x\tilde{s}(p,x)}.
\end{equation}
Since low $p_t$ particles spend more time in the fireball, and since the decay rate falls as $1/E$, escape rates are small at low $p_t$.

The corresponding classical calculations were performed by solving for the trajectories of particles through the mean field and are shown alongside the corresponding quantum calculations in Fig. \ref{fig:qclass}. The apparent sideward source sizes from the calculations used for Fig. \ref{fig:qclass} agree well with one another, differing by only a percent or two for $p_t>100$ MeV/$c$. Even at $p_t=40$ MeV/$c$, the calculations agree to better than 10\%. As explained above, the agreement should be expected to worsen for smaller $a$. Repeating the calculations for $a=0.5$ fm, discrepancies increased to the level of a few percent for $p_t>100$ MeV/$c$, and notably higher for $p_t\sim 50$ MeV/c. However, it is hard to physically motivate such a sudden change in densities.

One can also understand the connection between classical and eikonal approaches such as those discussed in Ref.s \cite{Kapusta:2005pt,Wong:2004gm,Chu:1994de}.
Since eikonal phases appear in interferences as $\delta(p_1,x)-\delta(p_2,x)$, and since the derivatives of phase shifts can be identified as the spatial offset incurred by slowing down or speeding up in a classical path through the potential, an connection between eikonal and classical approaches ensues for $p_1\sim p_2$ \cite{Pratt:2005hn}.

\begin{figure}
\centerline{\includegraphics[width=0.35\textwidth]{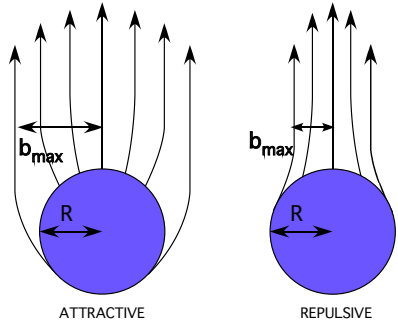}}
\caption{\label{fig:cartoon}
Trajectories that lead to the same asymptotic momentum are altered by an attractive mean field (left). The phase space distribution for particles of this given momentum are widened in coordinate space by a factor $b_{\rm max}/R$. Repulsive mean fields (right) reduce the size of the region of the outgoing particles.
}
\end{figure}

In order to better illustrate the physics of the refractive distortion we consider a simple example where pions are emitted from the surface of a cylinder of radius $R$, where the in-medium mass of the pion at the surface is $m_{\rm med}$ and the asymptotic vacuum mass is $m_{\rm vac}$. We assume the mass returns to its vacuum value exponentially,
\begin{equation}
m^2(r)=m_{\rm vac}^2+(m_{\rm med}^2-m_{\rm vac}^2)e^{-(r-R)/a}.
\end{equation}
From time-reversal arguments a thermal source will emit particles of given momentum with the same trajectories as those that describe absorption. The asymptotic trajectories of particles with momentum $p_x=p_t$, $p_y=0$, that intersect with the cylinder have a uniform distribution of impact parameters up to a maximum $b$ as illustrated in Fig.~\ref{fig:cartoon}. The effect of an attractive mean field is to stretch the width of the phase space cloud by a factor $b_{\rm max}/R$. The sideward dimension measured in two-particle correlations is stretched by this factor.

The maximum impact parameter for capture can be found by combining conservation of energy and angular momentum, $L=bp_t$,
\begin{eqnarray}
\label{eq:econs}
p_t^2&=&p_r^2+V_{\rm eff}(r)\\
\nonumber
V_{\rm eff}(r)&=&\frac{b^2p_t^2}{r^2}+m^2(r)-m_{\rm vac}^2.
\end{eqnarray}
One can analytically solve for the maximum impact parameter $b_{\rm max}$ from which one can overcome the potential barrier with a give $p_t$. The high $p_t$ behavior is independent of $a$, and in the case where $a>R/2$, the result is independent of $a$ for all $p_t$.

\begin{figure}
\centerline{\includegraphics[width=0.45\textwidth]{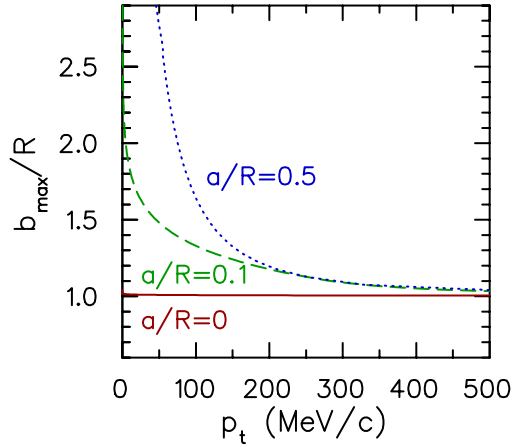}}
\caption{\label{fig:cylinder}
Assuming a static cylindrical source, the distortion to the sideward dimension due to the mean field is shown for attractive scalar fields. The field lowers the pion mass to 50 MeV/$c^2$ and falls off exponentially outside the emitting radius $R$ with a distance scale $a$. The distortion is stronger for larger ranges $a$ and at lower $p_t$.}
\end{figure}

Figure \ref{fig:cylinder} shows the ratio $b_{\rm max}/R$ as a function of $p_t$ for several values of $a$ given the case of a lighter in-medium pion mass, $m_{\rm med}=50$ MeV/$c^2$. For $a=0$, the mean field has no effect as the capture cross section does not extend beyond the cylinder. For the saturating value, $a=R/2$, the cross section is significantly enhanced. 
For $a=R/10$, the enhancement is identical to the saturating value at large $p_t$ and is somewhat reduced at low $p_t$. The enhancement of the apparent sideward size is characteristic of an attractive mean field. This is consistent with Liouvilles theorem, which states that contraction of the phase space density in momentum space as it leaves the attractive mean field should be accompanied by a growth of the phase space cloud in coordinates space.

The simple cylindrical picture was expanded to include both longitudinal and collective flow \cite{Pratt:2005hn}. For emission during the expansion stage, distortions were increased in magnitude since the boundary layer expands behind the pions, staying in the region of non-zero mean field for a longer time. Pions emitted during the collapse of the breakup surface were less affected since they spent less time near the surface.

The calculations presented here were only meant to be illustrative. The potential magnitude of mean-field effects and the ability of classical pictures to model the effects underscore the importance of incorporating mean field effects into hadronic Boltzmann descriptions of the breakup stage. Such calculations have been undertaken in the past, especially for lower-energy Fermi-velocity collisions, where the mean field is paramount \cite{gong,Bauer:1993wq}. In this energy range, phase space points used to generate correlations are typically taken from when a particle leaves the region of mean field, rather than when they have their last collision. Given the likelihood that the dispersion relations for pions are non-trivial, it is important that calculations incorporate momentum-dependent interactions. Such an effort is crucially important for reconstructing the space-time evolution of the fireball, and for understanding the pressure in the hadronic phase.

\begin{theacknowledgments}
Support was provided by the U.S. Department of Energy, Grant No. DE-FG02-03ER41259.
\end{theacknowledgments}

\end{document}